\documentclass[12pt]{article}
\usepackage{graphicx}
\newcommand{\sss}{\scriptscriptstyle}

\newcommand {\be}{\begin{equation}} 
\newcommand{\ee}{\end{equation}}    

\def\dds1{\frac{\partial}{\partial s_1}}

\def\d{d\kern-0.8 ex\vrule height 1.3 ex depth-1.24 ex width 0.7 ex
\kern 0.15 ex}
\def\D{D\kern-1.7 ex\vrule height .87 ex depth-0.8 ex width 0.7 ex
\kern 0.95 ex}

\begin{document}
%
%

\baselineskip 20 pt

\begin{center}

\Large{\bf Growing drift-cyclotron modes in the hot solar atmosphere}

\end{center}

\vspace{0.7cm}

\begin{center}

 J. Vranjes and S. Poedts

{\em Center for Plasma Astrophysics, Celestijnenlaan 200B, 3001 Leuven,
 Belgium.}

\vspace{5mm}

\end{center}

\vspace{2cm}

{\bf Abstract:}
Well-known analytical results dealing with ion cyclotron and drift waves and which follow from the kinetic theory are used and the dispersion equation, which describes coupled two modes, is solved numerically. The numerical results obtained by using the values for the plasma density, magnetic field and temperature applicable
to the solar corona clearly show the coupling and the instability (growing) of the two modes. The coupling happens at
very short wavelengths, that are of the order of the ion gyro radius, and for characteristic scale lengths of the
equilibrium density that are altitude dependent and may become  of the order of only a few meters. The demonstrated instability of the two coupled modes (driven by the equilibrium density gradient) is obtained by
using a rigorous kinetic theory model and for realistic parameter values. The physical mechanism which is behind the
coupling is simple and is expected to take place throughout the solar atmosphere and the solar wind which contain a
variety of very elongated density structures of various sizes. The mode grows on  account of the density gradient,
it is essentially an ion mode, and its further dissipation should  result in an increased ion heating.

\vfill
\pagebreak

\section{Introduction}

 The ion cyclotron resonance  and the ion cyclotron  mode have  been discussed   in the context of  problems related to the heating of
  the solar corona, e.g.,  Marsch et al. (1892), Cranmer et al. (1999), Cranmer (2000),  Tu $\&$ Marsch (2001), Markovskii (2001), Isenberg (2001),  Hollweg $\&$ Isenberg (2002).   This is due to various
reasons, such as  the evidence obtained from in situ measurements in the solar wind and coronal holes of resonant ion
cyclotron heating  (Hollweg $\&$ Isenberg 2002), and a preferential heating of coronal ions (with respect to electrons) which is most
dominant in the direction perpendicular to the magnetic field lines. The damping of such IC waves is believed to be a
good candidate for the consequent coronal heating and solar wind acceleration Markovskii (2001). However,  as in many
wave-heating scenarios of the corona proposed in the past, there is usually the problem of the source for the required
massive generation of such IC waves. In the literature, various effects have been proposed as sources for the IC
mode, like currents (Forslund 1970, Toichi 1971), global resonant MHD modes Markovskii (2001), etc. Yet, we note that these
effects themselves need some source, so the problem is not solved but merely shifted to another problem.
   The
 multi-fluid description of the  beam-driven  IC  wave  excitation is presented in Mecheri $\&$ Marsch (2007), where a tiny ion-beam  population is assumed to originate from small scale reconnections.  A rather large growth rate (of the order of $0.1 \Omega_p$, where $\Omega_p$ is the proton gyro-frequency) is obtained for relatively large values of the wave number, and for angles of propagation below $60^{\circ}$.

In one of the recent studies dealing with the generation of  ion cyclotron waves in the solar atmosphere Markovskii (2001)
it is pointed out that the solar atmosphere contains density inhomogeneities of various scales with density gradients
in the direction perpendicular to the magnetic field. Indeed, detailed studies performed in the past, which include
direct observations, reveal the existence of ray-like structures that span from very large perpendicular cross
sections, like in the case of polar plumes and streamers, down to very fine filamentary structures with a cross
section of the order of a kilometer. More details on that issue are given in   Woo $\&$  Habbal (1997), Woo
(1996), November $\&$  Koutchmy (1996), Karovska $\&$ Habbal (1991), and in references cited therein.

Fine filaments are visible even from ground-based observations like those during the eclipse in 1991 (November  and  Koutchmy  1997),
showing a slow radial enlargement of the structures (i.e., in the direction out of the Sun) which is consistent with
a low-beta plasma. In situ measurements by Voyager~1 and Voyager~2  (Woo $\&$  Habbal 1997), show that  the finest structures in
the slow solar wind at around $9\;R_{\odot}$ are about 3 times finer than those in the fast wind. Assuming a radial
expansion, they conclude that the transverse sizes of these highly elongated structures at the Sun are below 1$\;$km.
In   Karovska $\&$  Habbal (1991) an image restoration  method is used to study the structures of the quiet
Sun, with a spatial resolution of $5"$ and a temporal resolution of 5.5 minutes. The contour maps presented in the
work reveal the existence of numerous structures of various sizes.

The presence of density gradients in the direction perpendicular to the magnetic field lines implies the possibility
of  the existence of a  drift wave, a mode with the unique feature of being unstable both in the kinetic and the
fluid domain (hence the term {\em universally unstable mode}). Within  kinetic theory, the drift mode is unstable
even for a Maxwellian electron distribution. This is due to the inverse electron Landau damping effect. Within
fluid theory, the mode is also unstable, due to the common effect of the electron collisions and the ion inertia, in
the presence of a background density gradient.  A comparison between the kinetic and the fluid instability
(Goldston $\&$ Rutherford 1995) reveals that the resistive one is dominant provided that the electron parallel mean-free path is smaller
than the parallel wavelength. This is essentially the reason for the interest in the drift modes with very large
parallel wavelengths and relatively short perpendicular wavelengths.

 One particularly interesting feature of the
drift wave is related to its intrinsic nonlinearity (Hasegawa $\&$ Sato 1989).
The nonlinear terms, which follow from the convective derivatives in the momentum equations, are comparable to the linear ones provided that $k_\bot^2\rho_s^2 e\phi/(\kappa T_e)$ is of the same order as $\rho_s/L_n$. Here,
$\phi$ is the electrostatic wave potential, $\rho_s=c_s/\Omega_i$, $\Omega_i$ is the ion gyro-frequency,   $c_s^2=\kappa T_e/m_i$,
$L_n$ is the scale length of the density gradient,  and $\rho_s/L_n$ is usually a small quantity.  Hence, the mode becomes nonlinear even for very small perturbations provided that $k_\bot \rho_s$ is not much larger than unity.
Nevertheless, in the literature dealing with the waves and instabilities in solar plasma, the effects related to the
density gradient and the consequent drift waves are usually disregarded.

In our recent publications, some  aspects of the drift wave instability  in the solar plasma have been discussed.  The collisional coupling between the drift and kinetic Alfv\'en waves in the upper solar atmosphere (Vranjes $\&$ Poedts 2006)  reveals a strongly  growing drift mode in the chromospheric plasma. The similar collisional instability of the mode, though with a much  smaller increment, has been demonstrated also for the coronal plasma. In both cases, the kinetic Alfv\'en part of the mode is shown to be collisionally damped. In the presence of
a plasma flow along the magnetic field lines the drift mode is subject to a reactive-type instability, provided that
this background velocity has a gradient in the perpendicular direction. Such a problem has been discussed in our
recent work (Saleem, Vranjes $\&$ Poedts 2007) dealing with solar spicules. The drift mode has been shown to be unstable for typical spicule
characteristic lengths of the density and the shear flow gradients, i.e.\ in the range of a few hundred meters up to
a few kilometers, yielding wave frequencies of the order of a few Herz.

One of the basic properties  of the drift mode  is the low frequency
$|\partial/\partial t|\ll \Omega_i$.  In the case when this  low-frequency limit is not well satisfied, there appears a coupling between the drift and ion-cyclotron modes. In the limit when  $|\partial/\partial t|\sim \Omega_i$,  this  coupling is very effective and well known (Ichimaru 1973). In the simplest case, it yields a coupled unstable drift-cyclotron  mode with the
instability driven by the plasma density gradient. The first demonstration of this instability was  performed by
  Mikhailovskii $\&$  Timofeev (1963).

To the best of our knowledge, the   perpendicular density gradients, and the coupling of the
drift and IC modes in the regime $\omega \sim \Omega_i$ together with the consequent wave instability, have never
been discussed in the context of the heating of the solar corona and the acceleration of the solar wind. In the
present work we  describe the basic instability of the mode  to  focuss attention   on  the
existence of such an instability that may be widespread and may contribute substantially to the problems discussed
above. The term 'fine structures',  used in the text above for the structures observed in the solar corona still  denotes spatial scales that are much larger than those used in our  present study, in Sections 2 and 3. In other words, there is no  observational support for the existence of such tiny density inhomogeneities, and it is not expected in the near future. Yet, in view of the  variety of density structures at larger and observable spatial scales, and having no obvious physical effect that would prevent inhomogeneities at  scales that are even shorter than those observed so far,  such micro scale plasma inhomogeneities cannot be excluded.

\section{Drift-cyclotron instability}

Following Ichimaru (1973), the plasma dielectric function in the case of a negligible parallel wave vector and for
the frequency limit $\omega \sim \Omega_i$, and $k_{\bot}=k_y=k$ within the kinetic theory for hot ions and
electrons, is given by (in the local approximation)
\[
\epsilon(k, 0, \omega)= 1+ \frac{k_e^2}{k^2} \left[\frac{\omega_{*e}}{\omega} + 1 - \Lambda_0(\beta_e)\right]
\]
\be
+ \frac{1}{k^2 \lambda_{\sss D}^2} \left\{ 1- (\omega-\omega_{*})\left[ \frac{\Lambda_0(\beta)}{\omega} +
\frac{\Lambda_1(\beta)}{\omega-\Omega_i}\right]\right\}. \label{e3} \ee Here, \be
\beta_e= k^2\frac{T_e}{T_i}\frac{m_e}{m_i}\rho_{\sss L}^2 < 1, \quad \beta = k^2 \rho_{\sss L}^2 \gg 1,  \label{e4}
\ee
\[
\omega_*=\omega_{*e}T_i/T_e, \quad \omega_{*e}= \frac{n_0'}{n_0}\frac{T_e k_y}{m_e \Omega_e}, \quad k_e=\frac{\omega_{pe}}{v_{{\sss T}e}},
\]
\[
\Lambda_n(X)\equiv I_n(X) \exp(- X), \quad \rho_{\sss L}=\frac{v_{{\sss T}i}}{\Omega_i},
\]
$I_n$ denotes the modified Bessel function of the $n-$th order,  and the prime denotes the derivative in the direction perpendicular to both the wave-vector and the magnetic field. Using $\Lambda_n(\beta)\rightarrow (2 \pi
\beta)^{-1/2} \exp(-n^2/2\beta)$ (for $\beta\rightarrow \infty$), we have $\Lambda_0(\beta)\simeq 1/[(2\pi)^{1/2} k
\rho_{\sss L}]\equiv \delta$. In view of the first expression in Eq.~(\ref{e4}), we have $\Lambda_0(\beta_e)\simeq 1-
\beta_e$, while from the second one we have $\delta\ll 1 $. The dispersion equation then becomes
\be
\left(1+ k^2 \lambda_d^2 - \delta\right) \omega^2 - \left[\Omega_i \left(1+ k^2 \lambda_d^2\right) + \omega_*(1-
\delta) \right] \omega+ \omega_*\Omega_i=0,  \label{e5} \ee
\[
\lambda_d^2=\rho_{\sss L}^2 \frac{m_e}{m_i} + \frac{1}{k_e^2}\frac{T_i}{T_e}.
\]
In the two limits ($\omega\ll \Omega_i$ and $\omega \sim\Omega_i$), the two modes are the drift wave and the IC wave,
respectively, $\omega_1=\omega_*/(1+ k^2 \lambda_d^2)$ and  $\omega_2\sim \Omega_i[1+ \delta/(1+ k^2 \lambda_d^2)]$.
The instability may appear at the point of eventual intersection of the two dispersion curves, and the instability
condition  reads:
\be 4 \omega_* \Omega_i  \left(1+ k^2 \lambda_d^2 - \delta\right)>\left[\Omega_i \left(1+ k^2 \lambda_d^2\right) +
\omega_*(1- \delta) \right]^2. \label{e6} \ee
Below, we apply these expressions to the solar atmosphere in order to see if there is  a window in the relevant
parameter domain allowing for the instability.

\section{Application to the solar atmosphere}

The necessary instability condition (\ref{e6}) can be satisfied for a chosen set of plasma parameters $n_0, T, B_0,
L_n=(n_0'/n_0)^{-1}$ and for a given wavelength. Because of the horizontal and vertical stratification, various
values may be considered for the density, the temperature and the magnetic field. Yet, physically, in order to
have an instability, the frequencies of the drift and IC modes must become close to each other, and this is most
easily controlled by the density inhomogeneity scale-length $L_n$ and/or the wave-length.
 Here,  $L_n$ is a local, spatially dependent parameter that  determines the
 local properties of the drift-cyclotron mode.  Assuming  a cylindric elongated density
 structure with a radius $r_0$ and  with a Gaussian radial
 density distribution
 \[
 n_0(r)=N_0 \exp(-r^2/a^2),
 \]
where $N_0$ is the density at the axis of the cylinder, and $a$
 determines the radial decrease of the density, we have $L_n(r)=n_0(r)/n_0'(r)= a^2/(2r)$. Hence, we have a
 radially changing scale-length, which  goes to
infinity at the  center  and decreases towards  the boundary. In the
 case of an e-folding decrease along $r$,  we have $a=r_0$,  $n_0(r_0)/N_0=37 \%$,
and $L_n/r_0=1/2$. For such a density profile $L_n$ is minimum in the outer region of the plasma column. As a result, in the eigen-mode analysis  of the drift wave in the cylindric geometry  the amplitude and the increment  of the drift  mode are maximum in the same region (Bellan 2006,  Vranjes $\&$ Poedts 2005).

Hence, Eq.~(\ref{e5}) is
solved numerically in terms of the wavelength and the density scale length,  for parameter values applicable to the solar corona.
As an example we take $B_0= 10^{-3}\;$T, $n_0= 10^{13}\;$m$^{-3}$, $T_e=T_i=10^6\;$K,  that may be used to describe the physical properties of the plasma at the altitude of around one solar radius,   and  we have chosen a very
short density inhomogeneity scale-length, \, $L_n=5\;$m.  In the case of the Gaussian profile discussed above and for the e-folding decrease,  this yields
the characteristic radius of the structure $r_0\simeq 10$ m.   Such small values for $L_n$ are necessary to obtain  high
values for the drift wave frequency because it is proportional to $1/L_n$. For these parameters, the plasma beta is
$0.00035$ and we have a proper electrostatic limit. The result is presented in Fig.~1.  The instability develops in a
narrow range of wave-lengths  around  $1\;$m. Note that the ion gyro radius for these parameters is equal to
$\rho_{\sss L}=0.94\;$m. For larger wavelengths, the frequencies of the two modes become well separated and the
instability vanishes. The maximum increment $\omega_i$ is $\approx 5250\;$Hz at the real frequency
$\omega_r=102830\;$Hz. This value of the increment is lower than the approximate theoretical value (Mikhailovskii $\&$ Timofeev 1963) given
by $\Omega_i (m_e/m_i)^{1/4}$.  The small values for $L_n$ imply a relatively short time for the existence of such structures, making them difficult to detect. As  seen from Fig.~1, the frequencies of the corresponding modes are high, of the order of $10^5$ Hz. On the other hand the ion collision frequency $ \nu_{ii}= 4 n_{i0} (\pi/m_i)^{1/2} [e_i^2/(4 \pi
\varepsilon_0)]^2 L_{ii}/[3 (\kappa T_i)^{3/2}]$ for the given parameters is about $10^{-2}$ Hz. The perpendicular ion diffusion coefficient  (Chen 1988) is $D_\bot\approx \kappa T_i \nu_i/(m_i \Omega_i^2)=0.01$ m$^2$/s. The diffusion velocity in the
 direction of the given density gradient is $D_\bot \nabla n/n=2$ mm/s only.  So we have
about 7 orders of magnitude difference for the two characteristic times, and this is enough time for the instability to develop before the equilibrium  density structure disappears.

In Fig.~2, we fix the wave-length at $\lambda=0.95\;$m, for the same parameter values as in Fig.~1, and calculate the
frequency in terms of the density length scale $L_n$. The maximum increment $\omega_i=5361\;$Hz is obtained at
$L_n=5.3\;$m, and the real part of the coupled mode frequency in that case is $\omega_r=102504\;$Hz.  Compared to the recent results of Mecheri $\&$ Marsch (2007), the present  growth-rate driven by the density gradient  appears to be smaller.

 We note that in some studies,  e.g., Coles and Harmon (1989), the existence of a wave-number cutoff has been found at the ion inertial length $\lambda_i=c/\omega_{pi}$, preventing density perturbations for wavelengths shorter than $\lambda_i$. However, this should not be confused with   the density structures in our work. We are dealing with electrostatic perturbations at considerably shorter time and space scales that are presently difficult to detect.  More importantly, there seems to be no definite consensus about the nature and the origin of the mentioned wave-length cutoff. Although it is a separate issue, in our view it seems very likely that the mentioned cutoff should be attributed
to the problem of electromagnetic perturbations  at spatial/time scales satisfying $(\omega/k)^2\ll c^2$, and for
 wavelengths below the ion inertial length. As it is known from plasma theory (Shukla et al. 2001; Stamper $\&$ Tidman 1973; Vranjes et al. 2007; Yu $\&$ Stenflo 1985) in this
case, the displacement current in the Amp\`{e}re law $
 \nabla\times B=\mu_0 \vec j+ \mu_0 \varepsilon_0 \partial \vec
 E/\partial t$ can be omitted,  and the ion perturbations become  negligible.  Further, setting   Amp\`{e}re's law (without the
last term) in the electron continuity equation reveals that the
 electron density perturbations vanish.

\section{Conclusions}

The instability discussed here implies very short scale lengths for the inhomogeneity of the equilibrium density
and/or a  very weak magnetic field. Only in such circumstances can the frequencies of the drift and IC modes  become
close to each other so that the two modes can effectively couple. Short density scale lengths are presently not
directly observable in the solar atmosphere. However, we have learned   that  improvement of the
resolution in observations of phenomena in the solar atmosphere tends to result in an increased variety and
complexity of the density and/or magnetic field structures at short scales, showing very tiny filaments at scales
below 1~km. Therefore, one may expect a larger variety at even shorter scales. The present analysis clearly
demonstrates the instability of perpendicularly propagating modes at frequencies in the range of the ion cyclotron
frequency and at wavelengths of the order of the ion gyro radius. The tiny filaments are very
elongated, they may extend to many solar radii, and the growth of the mode and the consequent dissipation
and heating of ions  may take place over  large distances. The numbers used here are for the solar corona however,
the large radial (from  the Sun) length of the structures and the consequent decreasing of the magnetic field
intensity implies larger density scale lengths at which the instability takes place. This can be easily shown by
reducing the magnetic field to $10^{-4}\;$T and the number density by one order of magnitude. As a result, the
necessary density scale length $L_n$ for the unstable modes becomes of the order of 60 meters.  Therefore the
development of unstable growing modes may take place at large distances along the same density filaments that
pervade the corona and spread within the solar wind. The  presence of hotter ions in and around such filamentary
structures should be interpreted as an indication and a signature of the instability.

There have  been  intensive searches for possible mechanisms for the excitation of such modes. Our
analysis  is based on one  such mechanism that is well known, but  not used in the context of the solar plasma, and it requires a step  beyond the widely used MHD model. At higher  densities or temperatures the plasma beta becomes higher so that electromagnetic effects should
be included. However, it is known that eventual bending of the magnetic field lines implies an additional obstacle for
 electron motion along the magnetic lines, making the drift-type modes even more unstable (Vranjes et al. 2007b). The parameter
values that we use here are realistic and the instability that has been demonstrated for the given cases is thus
physically very likely.

Acknowledgements:

These results are  obtained in the framework of the projects G.0304.07 (FWO-Vlaanderen), C~90205 (Prodex),
GOA/2004/01 (K.U.Leuven),  and the Interuniversity Attraction Poles Programme - Belgian State - Belgian Science
Policy.



\vfill

\pagebreak

\noindent {\bf Figure captions:}

\begin{description}
\item{1.} The frequencies and the increment (multiplied by 10) of the coupled drift-cyclotron mode in terms of the wave-length.
\item{2.} The frequencies and the increment of the coupled drift-cyclotron mode in terms of the characteristic density gradient length. Here, the increment is multiplied by 10.
\end{description}

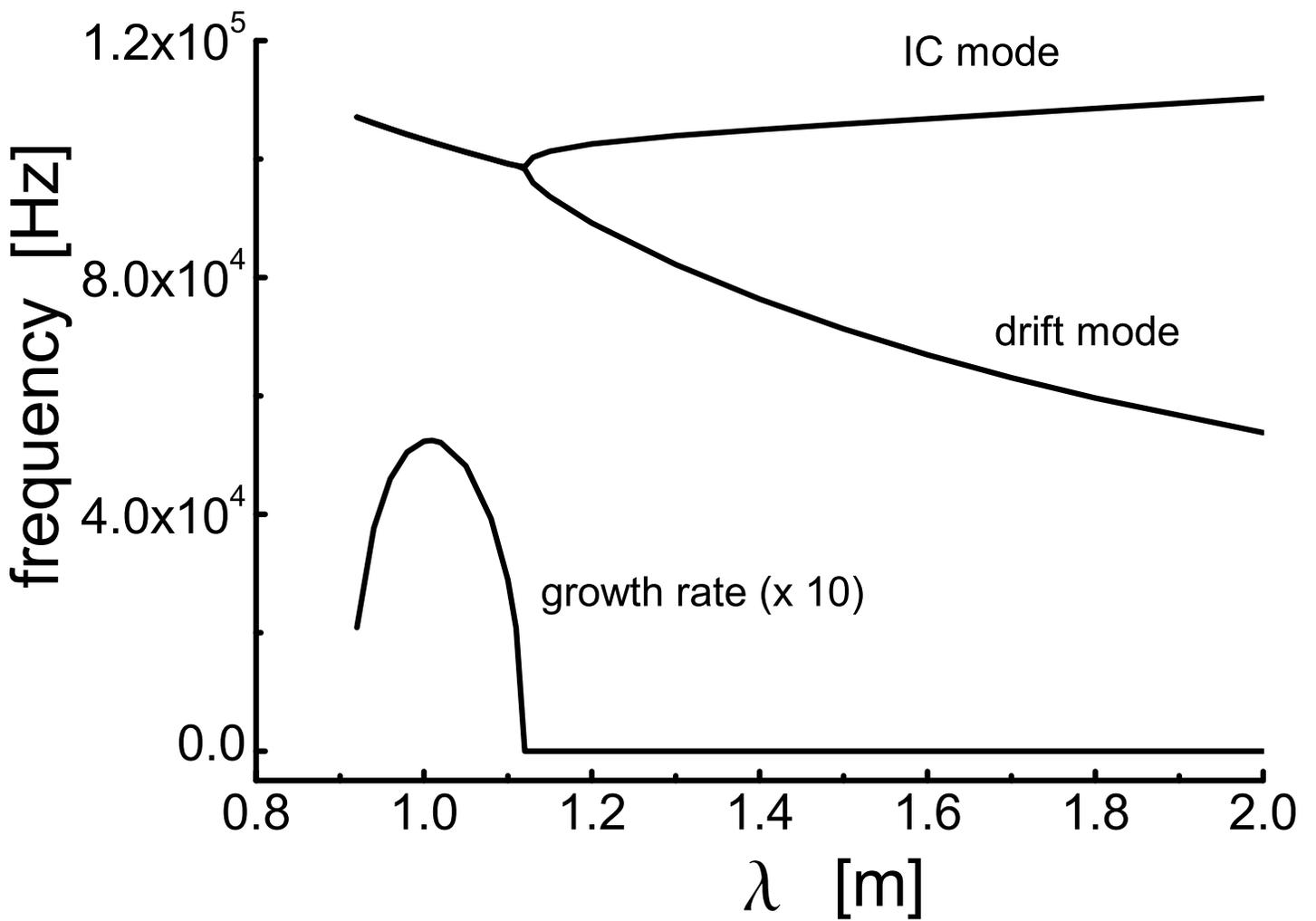
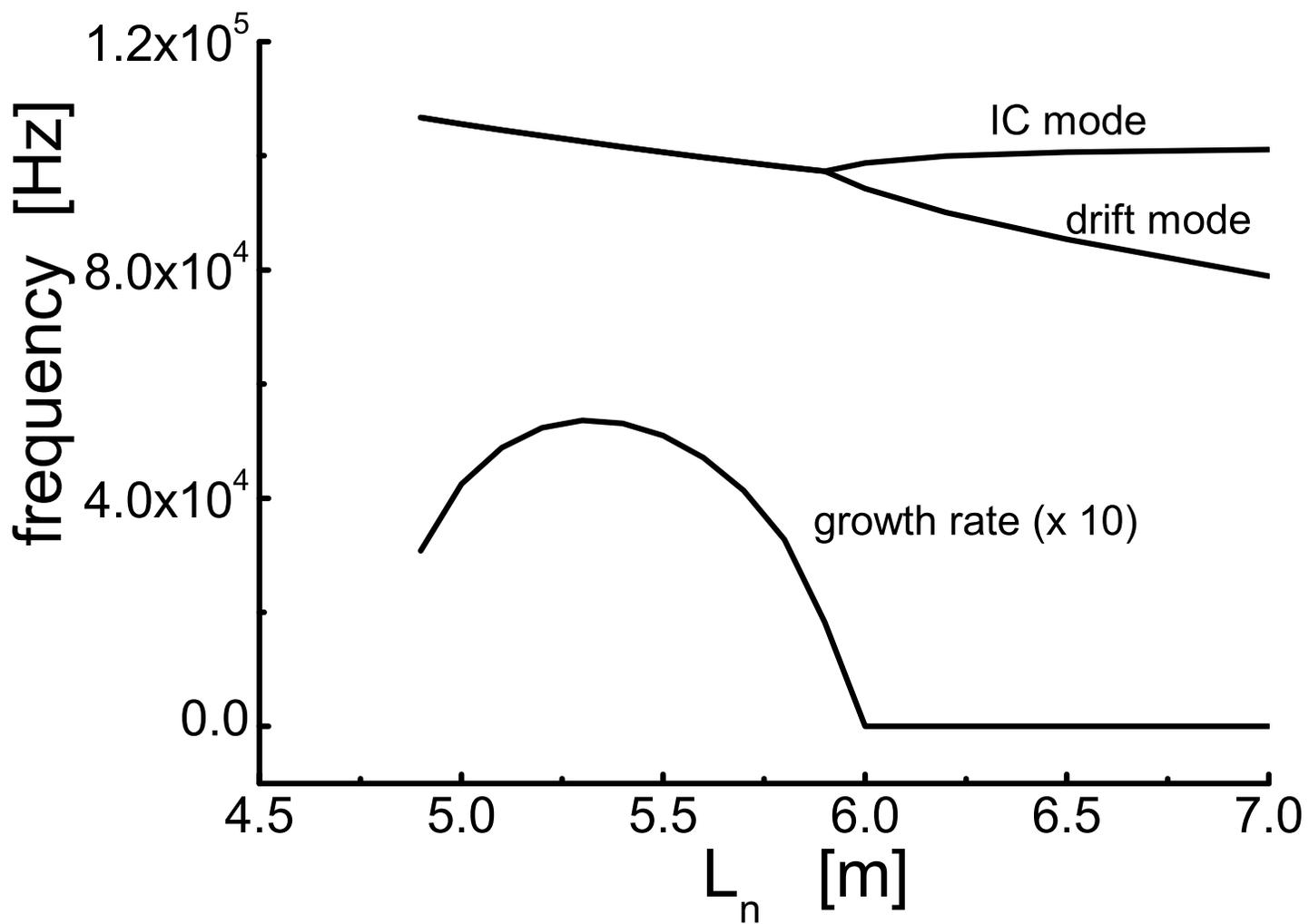
\end{document}